\documentclass{article}

\usepackage{arxiv}

\usepackage[utf8]{inputenc} 
\usepackage[T1]{fontenc}    
\usepackage{hyperref}       
\usepackage{url}            
\usepackage{booktabs}       
\usepackage{amsfonts}       
\usepackage{nicefrac}       
\usepackage{microtype}      
\usepackage{lipsum}
\usepackage{color}
\usepackage{algorithm}
\usepackage{algorithmic}
\usepackage{amsmath,amsfonts,amssymb}
\usepackage{subfigure}
\usepackage{lineno}
\usepackage{graphicx}
\usepackage{bm}
\usepackage{subfigure}

\title{
Machine-learning-assisted phase-amplitude reduction for fast synchronization of airfoil wakes \\with constrained fluctuations
}

\author{
Daiki Beppu$^{[1]}$, Vedasri Godavarthi$^{[2]}$, Koichiro Yawata$^{[3]}$, Hiroya Nakao$^{[3]}$, and Kai Fukami$^{[1,*]}$
\\
\\
$[1]$ Department of Aerospace Engineering, Graduate School of Engineering, Tohoku University, Sendai, 980-8579, Japan\\
$[2]$ Department of Mechanical Engineering, Johns Hopkins University, Baltimore, MD, 21218, USA\\
$[3]$ Department of Systems and Control Engineering, Institute of Science Tokyo, Tokyo, 152-8552, Japan\\
$^{*}$Corresponding author: kfukami1@tohoku.ac.jp
}

\begin{document}
\maketitle

\begin{abstract}

\vspace{-2mm}
    This study considers rapidly modifying the wake shedding frequency of the flow around an airfoil using sparse sensor information, subject to constraints on the lift coefficient fluctuations. 
    This is achieved by combining phase-amplitude reduction with nonlinear machine-learning-based sparse sensor reconstruction. 
    We derive time-varying phase and amplitude sensitivity fields that identify the optimal spatial locations and timing for actuation from merely three sensors. Through the sensitivity fields, we analytically obtain the optimal waveform for fast synchronization of wake shedding frequency while minimizing amplitude deviation of aerodynamic responses.
    The proposed approach is evaluated using flows over various NACA airfoils at several post-stall angles of attack, all of which exhibit unsteady periodic vortex shedding.
    With the identified optimal forcing, the wake frequency is altered eight times faster than with a standard sinusoidal actuation.
    Furthermore, the amplitude-penalized forcing achieves $20\%$ suppression of the lift coefficient fluctuation compared to the optimal forcing without amplitude penalty.
    The current amplitude-penalized technique may offer an efficient path for fast flow modification without causing detrimental fluctuations in periodic aerodynamic and aeroelastic systems with fluid-structure interactions.
\end{abstract}

\section{Introduction}
\label{sec:intro}
\vspace{-2mm}

{
Manipulating fluid flows is challenging due to their high dimensionality and multi-scale nature. When the underlying physics exhibits recurring patterns such as periodicity, the potential for simplification through low-dimensional modeling arises, making control strategies more tractable~\cite{taira2017,brunton2022data,Fukami_JFMPerspective}.
For such low-dimensional systems, it is often possible to recover the spatiotemporal behavior of unsteady flows from sparse sensor information~\cite{fukami_fukagata_taira2023,erichson2020}. 
This study uses phase-amplitude low-order modeling and machine-learning-based sparse sensor reconstruction to achieve fast frequency modification of unsteady aerodynamic flows.
}

{
Rhythmic phenomena are ubiquitous in various fields, including biology~\cite{Winfree1967, ermentrout2010mathematical}, chemistry~\cite{Kuramoto1984}, and fluid dynamics~\cite{Taira2026}.
Flow instabilities can result in oscillatory behaviors, such as K\'{a}rm\'{a}n vortex shedding~\cite{roshko1954,bearman1984,williamson2004}, flutter~\cite{amandolese2013,tang2003,godavarthi2026energy}, and buffet instability~\cite{giannelis2017review,Fukami_Iwatani_Maejima_Asada_Kawai_2025}, which possess the characteristics of limit-cycle oscillations: returning to their original state after perturbation.
Analyzing and controlling these phenomena is critical for engineering applications, enabling noise mitigation, enhancing aerodynamic performance, and energy efficiency in energy harvesting applications. 
However, establishing control strategies remains a significant challenge even for such periodic flows~\cite{colonius2011}, due to inherent nonlinearity and time-varying base states~\cite{linot2024, ashtiani2025}. 
}

{
Floquet analysis has been traditionally used for the stability analysis of periodic systems.
These systems are linearized about a periodic solution, and the perturbation growth is predicted using Floquet multipliers.
Recently, a complementary phase reduction approach ~\cite{Kuramoto1984, Winfree1967, Nakao2014} has been applied to periodic fluid flows to identify the effective timing for actuation. This approach describes periodic phenomena through a single scalar phase variable. By decreasing the dimensionality of the system, this formulation facilitates the simplification of high-dimensional perturbation dynamics. Equipped with a phase-based model, developing a control strategy becomes more tractable, which is evident from a range of studies.
Examples include Hele-Shaw flows~\cite{Kawamura2013, KAWAMURA201511}, flow around a cylinder~\cite{Taira2018, Khodkar2020,Khodkar2021, Loe2021, Kim2024}, a flat plate~\cite{Iima2021, Iima2024}, an airfoil~\cite{Nair2021, Kawamura2022, Godavarthi2023,Sumanasiri2025}, supersonic turbulent cavity flows~\cite{godavarthi2025phase}, and transonic airfoil buffet~\cite{candon2026phase}.
}

{
Effectively manipulating periodic behavior requires identifying an appropriate timing at which the phase of the system is sensitive to perturbations~\cite{asztalos2021modeling,beckers2024deep}. 
This temporal sensitivity can be quantified by the phase sensitivity function. 
There are two standard approaches for determining the sensitivity. 
One is the direct method, which observes the phase shift in terms of advancement or delay to small impulse perturbations to the system~\cite{Nakao2016, Taira2018}. 
Another is the adjoint method, which calculates the spatial sensitivity fields with respect to all the flow variables considering adjoint dynamics~\cite{ermentrout2010mathematical}. 
When a numerical solver is available, the adjoint method not only provides the sensitive timing but also visualizes the whole spatial sensitivity field, enabling the identification of where and how an actuation is efficient~\cite{Kawamura2022}. 
Once the sensitivity profile is established, it is possible to derive the optimal waveform for frequency modification through the lens of synchronization~\cite{Zlotnik2013}. By maximizing how quickly the natural frequency synchronizes with the forcing frequency, a quick flow modification strategy can be determined~\cite{Godavarthi2023}.
}

{
Even when frequency modification through synchronization is achieved, the limit-cycle amplitude may deviate, resulting in detrimental fluctuations if the control strategy neglects amplitude dynamics. 
For aerodynamic applications, the phase and amplitude are often determined based on the lift coefficient~\cite{Taira2018, Godavarthi2023, Kim2024}, which is of critical importance to flight performance. 
In other words, performing the synchronization-based control without special care for amplitude may cause degradation of aerodynamic characteristics. 
Phase-reduction analysis has recently been extended to amplitude dynamics~\cite{mauroy2013,wilson2016,Shirasaka2017,Takata2021}. 
These phase and amplitude sensitivity functions are in fact the left Floquet eigenfunctions corresponding to the zero eigenvalue and other eigenvalues of the periodic system.
By leveraging the amplitude sensitivity, it becomes possible to modify the frequency while simultaneously controlling the amplitude of the limit-cycle oscillator~\cite{Takata2021}. 
The amplitude-constrained phase control has been shown to be effective for gust mitigation of periodic airfoil wakes~\cite{Fukami2024}.
}

{
Constructing amplitude sensitivities for a high-dimensional flow field is challenging as it involves the Floquet analysis. 
However, recent advances in machine learning have enabled phase-amplitude reduction in a low-order latent subspace~\cite{Yawata2024, Yawata2025, Fukami2024,ushida2026od}. 
Such a data-driven technique makes phase-amplitude dynamics relatively tractable. 
Since obtaining the full spatial sensitivity field is often computationally expensive~\cite{Kawamura2022}, it is ideal to perform the phase-amplitude-based analysis of fluid flows in a tractable manner by relying on measurable, low-order quantities such as sensor readings.
}

\begin{figure}
    \begin{center}
        \includegraphics[width=1\textwidth]{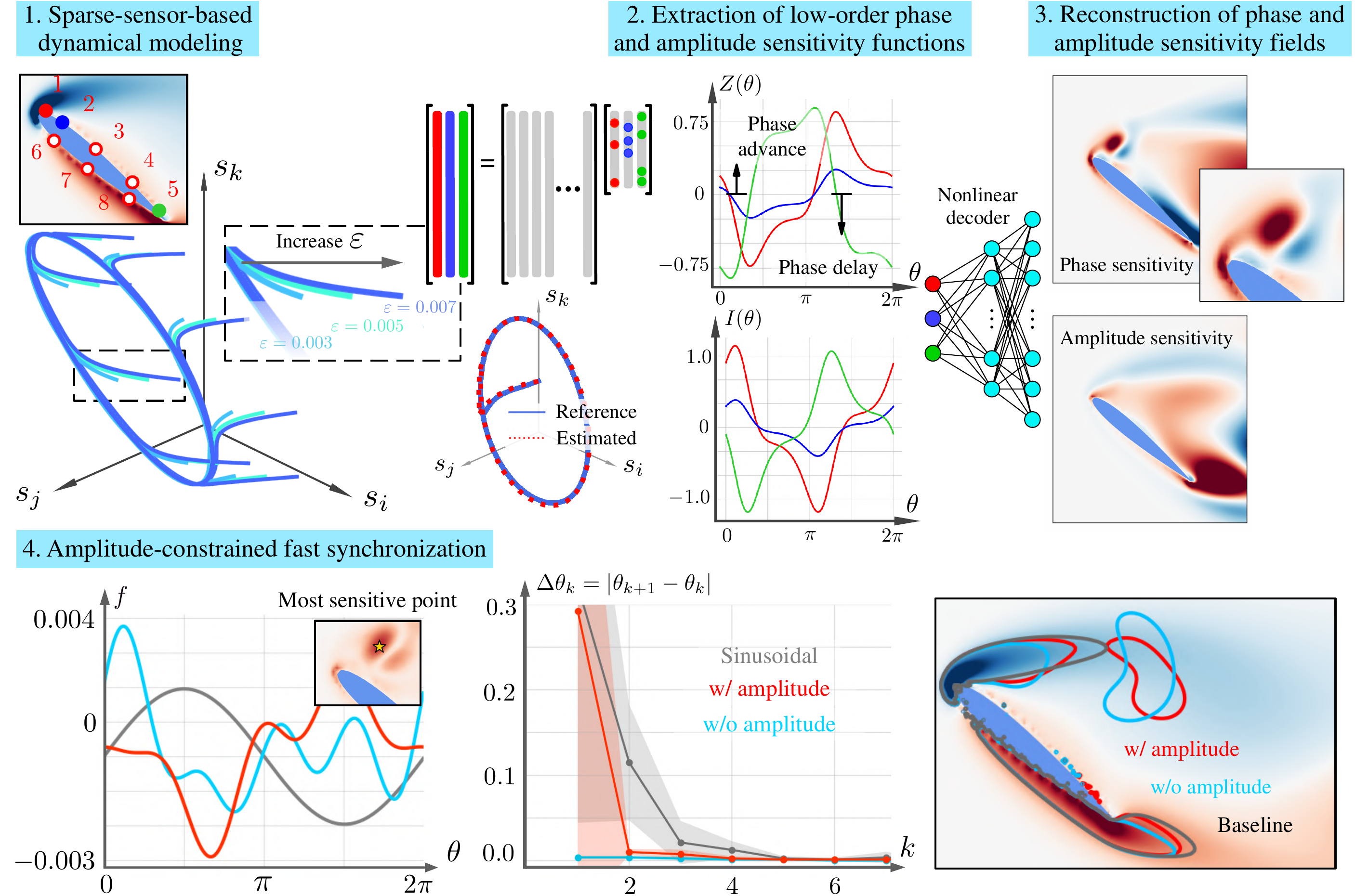}
    \end{center}
    
    \vspace{-2mm}
    \caption{
    Overview of the present study: 1. Sparse sensor-based dynamical modeling~(\S\ref{SINDy}). 2.~Extraction of low-order phase and amplitude sensitivity functions (\S\ref{phase reduction}).
    3.~Reconstruction of phase and amplitude sensitivity fields (\S\ref{decoder}).
    4.~Amplitude-constrained fast synchronization (\S \ref{optimal entrainment}).
    }
    \vspace{-3mm}
    \label{fig1}
\end{figure}

{
The present paper proposes a data-driven approach to achieve fast synchronization of the airfoil wake with the care of amplitude modification for aerodynamic coefficients.
This is achieved by a framework designed to approximate phase and amplitude sensitivity fields from sparse sensors. 
An overview of this study is presented in figure~\ref{fig1}.
We first develop a dynamical model in a low-order sensor subspace, and learn the dynamics in the vicinity of a limit cycle of periodic wake shedding.
The phase and amplitude sensitivity functions are then derived using sparse dynamical modeling and the adjoint analysis in the low-order space.
Assisted by a nonlinear decoder trained to reconstruct a high-dimensional vortical flow from sparse sensors, the high-dimensional spatial sensitivity fields are approximated from the resulting low-order sensitivity functions.
We then consider performing frequency-based flow control of wake shedding around an airfoil by analytically obtaining an optimal actuation waveform, achieving fast modification of wake frequency while suppressing the amplitude deviation of aerodynamic coefficients.

}

{
The present paper is organized as follows. 
Data preparation and aerodynamic flow physics, including parametric properties such as angle of attack and airfoil thickness, are introduced in \S \ref{sec2}. 
The framework for using the phase-amplitude reduction and nonlinear machine learning to obtain sensitivity fields from sensor measurements and control strategies is described in \S \ref{sec3}.
The results on the sensitivity fields and the control assessment are presented in \S \ref{sec4}. 
Conclusions are lastly remarked in \S \ref{sec5}.
}

\section{
Data curation and aerodynamic characteristics of post-stall airfoil wakes
}
\label{sec2}

{
This study aims to derive optimal control strategies for amplitude-penalized fast-frequency modification of periodic aerodynamic flows.
We propose a data-driven framework, composed of phase-amplitude reduction, a field decoder, and sparse dynamical modeling, as graphically summarized in figure~\ref{fig1}. 
This section first details the data preparation process, including the numerical solver and dataset specifications.
The unsteady periodic flow dynamics are subsequently examined.
}

{
We consider a two-dimensional incompressible laminar flow over different airfoil thicknesses and at different angles of attack, under a chord-based Reynolds number of $Re = u_{\infty}c/\nu=100$.
Here, $u_{\infty}, c$ and $ \nu$ are the free-stream velocity, the chord length, and the kinematic viscosity, respectively.
We prepare datasets for the current analysis by numerically solving the incompressible Navier--Stokes equations,
\begin{align}
    \nabla\cdot\bm{u}&=0\\
    \frac{\partial\bm{u}}{\partial t}+\bm{u}\cdot\nabla\bm{u}&=-\nabla p   +\frac{1}{Re}\nabla^{2}\bm{u} + \varepsilon\bm{f},
        \label{NS equation}
\end{align}
where $\bm{u}$ is the velocity vector field and $p$ denotes the pressure field. 
To measure perturbation dynamics for phase-amplitude reduction, which will be discussed in section~\ref{phase reduction}, we introduce impulsive forcing $\varepsilon\bm{f}$ as a body force satisfying
\begin{align}
    \varepsilon\bm{f}=\delta(\bm{x}-\bm{x}_f)\eta(t)\hat{\bm{e}},
\end{align}
where the amplitude $\varepsilon \ll 1$, the location of forcing is $\bm{x}_f$, and the forcing direction is $\bm{\hat{e}}$. The function $\eta(t)$ is a Gaussian impulse of the form such that
\begin{align}
\eta(t) =
\varepsilon \delta(t) \approx \dfrac{\varepsilon}{\sqrt{2\pi}\sigma} \exp \left[ -\dfrac{1}{2} \left( \dfrac{t-t_0}{\sigma} \right)^2 \right],
\end{align}
where $\sigma$ is the width of the impulse, which is set to $\sigma=0.01$, and $t_0$ is the time instant of actuation.
}

{
\begin{figure}
    \centering
    \includegraphics[width=1\textwidth]{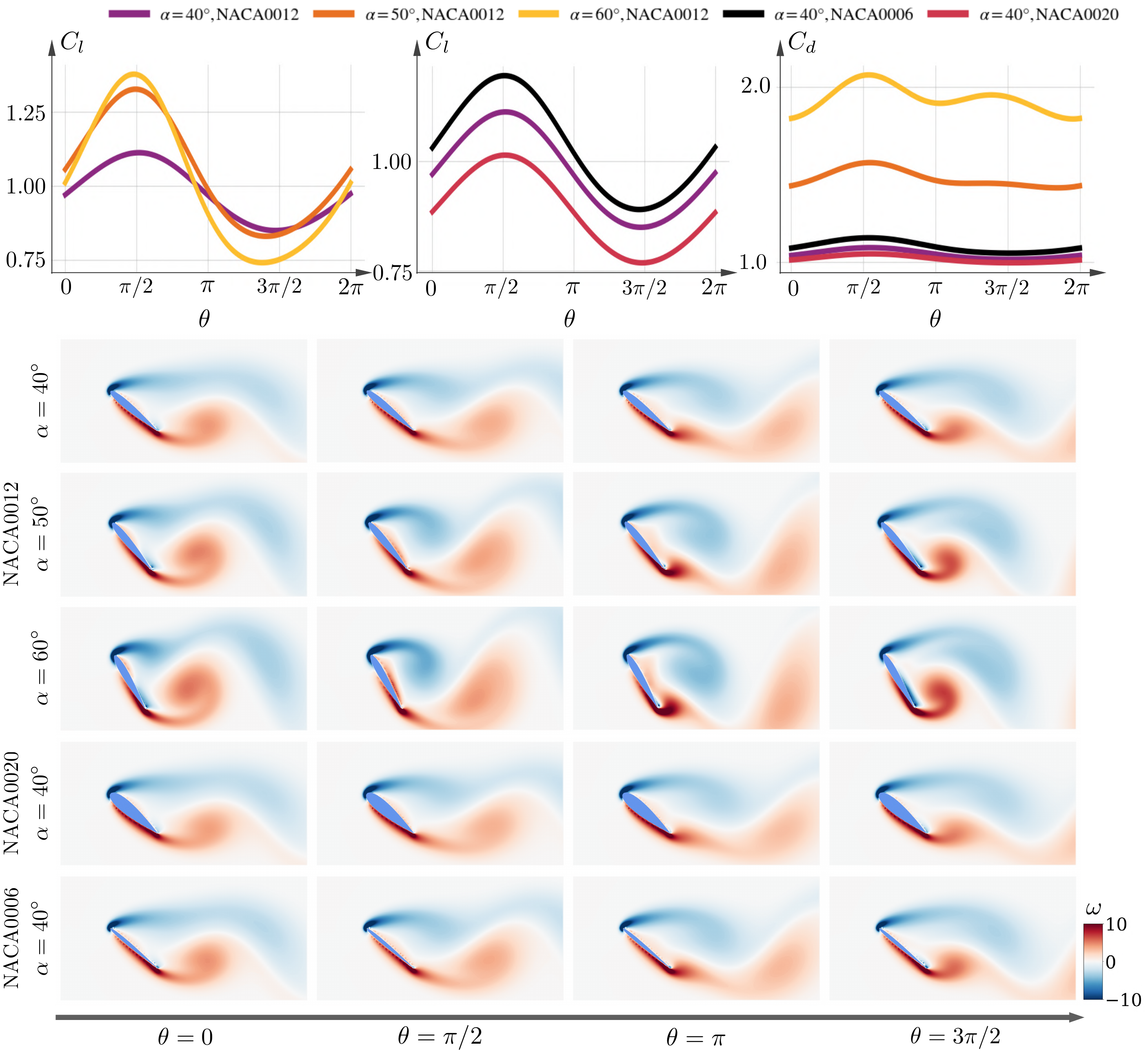}
    \vspace{-6mm}
    \caption{
    Aerodynamic coefficients and vorticity fields across various airfoil parameters.
    }
    \vspace{-4mm}
    \label{fig2}
\end{figure}
Direct numerical simulations with the fast immersed boundary projection method are used~\cite{Taira2007,Colonius2008}. 
This method uses a multi-domain Cartesian mesh, and the immersed body is represented as a diffused interface using boundary forces.
Three nested levels of the computational domain are used, with the finest level being $(x,y)/c \in [-1,6]\times[-1,6]$ and the largest domain being $(x,y)/c \in [-11.5, 16.5]\times[-11.5,16.5]$ in size, as the quarter-chord of the airfoil is placed at the origin.
The smallest mesh-grid size is set to $\Delta x_{\rm min}/c=y_{\rm min}/c=0.014$, and the time step is set to $\Delta t=0.005$. 
Verification and validation have been performed extensively by comparing with previous studies~\cite{DiIlio2018, Kawamura2022, fukami_taira_2023}. 
}

{
 Various parameter combinations, including angle of attack $\alpha\in\{40^\circ, 50^\circ, 60^\circ\}$ and airfoil thickness \{NACA0006, NACA0012, NACA0020\} are considered to assess the present data-driven technique. 
Let us exhibit in figure~\ref{fig2} the aerodynamic coefficients and vorticity fields across different parameters considered herein.
Here, the lift and drag coefficients are defined as $C_l=F_L/0.5\rho u_{\infty}^{2}$, and $C_d=F_D/0.5\rho u_{\infty}^{2}$, where $\rho$ is the free-stream density and $F_L$ and $F_D$ are the lift and drag forces, respectively.
At the current Reynolds number of 100, the flow is steady below $\alpha \approx 30^\circ$. The flow exhibits periodic vortex shedding for the considered angles of attack~\cite{Choi_Colonius_Williams_2015,Fukami2024}.
}

{
All the flows considered herein are post-stall, in which the vorticity fields exhibit separation at both leading and trailing edges, maintaining periodic vortex shedding. 
For these flows, the lift coefficient shows limit-cycle oscillations and the drag coefficient shows higher harmonics. 
Hence, $C_l$ can be treated as an observable to define the phase variable, $\theta$, on the $C_l-\dot{C_l}$ plane.
Here, $\theta = 0$ is chosen such that $C_l = [{C_l}]_t$ and $\theta=\pi/2$ corresponds to $C_l = {\rm max}~C_l$~\cite{Taira2018}, where $[\cdot]_t$ denotes the time-averaging operation.
}

{
The time-varying aerodynamic coefficients correspond to the periodic evolution of the vortical structures throughout an oscillation cycle. 
The formation of a leading-edge vortex induces strong suction due to enhanced circulation above the airfoil, yielding the maximum lift coefficient before the vortex sheds~\cite{Eldredge2018}.
At $\theta=\pi/2$, the negative vortex dominates the upper surface of the airfoil.
At this point, the bound circulation reaches its peak magnitude for the period.
In contrast, the positive vortex becomes dominant near the trailing edge at $\theta=3\pi/2$ while the suction-side leading-edge vortex detaches and convects away from the airfoil.
}

{
Although every case exhibits periodic behavior, the respective trajectory characteristics vary as the parameter changes.
The lift coefficient fluctuation deviates as the angle of attack increases, since the circulation of the leading- and trailing-edge vortices increases with angle of attack. This can be seen as the strengthened vortex roll-up at the leading and trailing edges.
The lift coefficient decreases as the airfoil thickness increases, which is attributed to the lower leading-edge curvature, thereby mitigating separation near the leading edge.
Regarding the drag coefficient, its dependence on thickness mirrors that of the lift coefficient, while exhibiting multi-frequency behavior.
}

\section{
Sensor-based phase-amplitude reduction and wake control framework
}
\label{sec3}

{
The flow physics considered above reveals that each flow scenario possesses periodic properties, suggesting the potential for low-order control through phase and amplitude dynamics.
We consider phase-amplitude reduction to characterize the high-dimensional perturbation dynamics about the periodic base state using the phase and amplitude variables. 
The phase variable, as considered above, captures the phase of the periodic vortex shedding. The amplitude variable captures the deviation of the perturbed trajectory from the baseline limit cycle.
}

{
Previous methods, such as a direct impulse-based method~\cite{Taira2018, Khodkar2020,Loe2021}, an adjoint-based approach~\cite{Kawamura2013, Kawamura2022}, and a Jacobian-free approach~\cite{Iima2021}, may be capable of determining sensitivity functions.
However, obtaining the full sensitivity fields has remained a challenge due to the difficulty of bridging low-dimensional phase-amplitude space and high-dimensional flow fields.
In response, the present study considers the integration of a nonlinear decoder and phase-amplitude reduction, enabling the identification of time-varying sensitivity fields from sparse sensor measurements.
We analyze and control periodic vortex shedding using phase-amplitude modeling through the following procedure:
\begin{enumerate}
    \item sparse dynamical modeling in sparse sensor space (\S ~\ref{SINDy})
    \item phase-amplitude reduction to derive phase- and amplitude sensitivity functions in sparse sensor space (\S ~\ref{phase reduction})
    \item reconstructing phase- and amplitude-sensitivity fields through nonlinear decoder (\S \ref{decoder})
    \item controlling aerodynamic flows with optimal waveform for fast synchronization (\S \ref{optimal entrainment})
\end{enumerate}
Through these steps, we identify phase- and amplitude sensitivity fields based on sparse sensor measurements. 
Subsequently, we design the optimal waveform that modifies the shedding frequency while achieving the suppression of lift-coefficient fluctuation. 
Hereafter, let us detail each step of the procedure above.
}

\subsection{Sparse identification of nonlinear dynamics}
\label{SINDy}

{
In the proposed formulation, we aim to reconstruct the sensitivity of the flow fields from that of the low-dimensional representation in the sensor space.
The first step to that end is to model the baseline and the perturbation dynamics of the low-order representation. 
Once appropriate sparse sensors are chosen, as will be discussed later in section~\ref{sec4},
we consider modeling the sensor space dynamics as ordinary differential equations (ODEs) via the sparse identification of nonlinear dynamics~\cite[SINDy;][]{Brunton2016}.
Given a low-dimensional state vector $\bm{\eta} \in \mathbb{R}^{n_{\mathbb{\eta}}}$, its dynamical system is written as
\begin{equation}
    \dot{\bm{\eta}}=\bm{F}(\bm{\eta})
    \label{sensor dynamics},
\end{equation}
where $\bm{F}$ represents the dynamics to be identified, and $\dot{\eta}$ is the time derivative of $\eta$. To determine the function~$\bm{F}$, discretized data in time are arranged into a data matrix $\bm{H}$.
The corresponding derivative matrix $\dot{\bm{H}}$ is evaluated using a second-order difference. Both these matrices are given as,
\begin{equation}
    \bm{H}=
    \left[
    \begin{array}{c}
        \bm{\eta}^T(t_1)\\
        \bm{\eta}^T(t_2)\\
        \vdots\\
        \bm{\eta}^T(t_m)\\
    \end{array} \right] =
    \left[
    \begin{array}{cccc}
         \eta_1(t_1)& \eta_2(t_1)&\dots& \eta_n(t_1)\\
         \eta_1(t_2)& \eta_2(t_2)&\dots& \eta_n(t_2)\\
         \vdots  &  \vdots  &\dots&\vdots   \\
         \eta_1(t_m)& \eta_2(t_m)&\dots& \eta_n(t_m)\\
    \end{array} \right] \in \mathbb{R}^{m \times n_{\mathbb{\eta}}},
    \label{sindy_s matrix}
\end{equation}

\begin{equation}
    \dot{\bm{H}}=
    \left[
    \begin{array}{c}
        \dot{\bm{\eta}}^T(t_1)\\
        \dot{\bm{\eta}}^T(t_2)\\
        \vdots\\
        \dot{\bm{\eta}}^T(t_m)\\
    \end{array} \right] =
    \left[
    \begin{array}{cccc}
         \dot{\eta_1}(t_1)& \dot{\eta_2}(t_1)&\dots& \dot{\eta_n}(t_1)\\
         \dot{\eta_1}(t_2)& \dot{\eta_2}(t_2)&\dots& \dot{\eta_n}(t_2)\\
         \vdots  &  \vdots  &\dots&\vdots   \\
         \dot{\eta_1}(t_m)& \dot{\eta_2}(t_m)&\dots& \dot{\eta_n}(t_m)\\
    \end{array} \right] \in \mathbb{R}^{m\times n_{\mathbb{\eta}}},
    \label{sindy_sdot matrix}
\end{equation}
where $m$ denotes the number of snapshots. To construct the right-hand side of equation~\ref{sensor dynamics}, a library matrix $\bm{\Phi}$, often consisting of nonlinear polynomial functions, is evaluated, where each term is referred to as a candidate for $\bm{F}$ as,
\begin{equation}
    \bm{\Phi}(\bm{\eta})=
    \left[
    \begin{array}{cccc}
     | & |  &  | &  |\\
     1 & \bm{\eta}  & \bm{\eta}^{P_2}  &  \bm{\eta}^{P_3}\\
     | & |  &  | &  |\\
    \end{array}
    \right]
    \label{sindy_library matrix},
\end{equation}
where quadratic and cubic polynomial terms are denoted as $\bm{\eta}^{P_2}, \bm{\eta}^{P_3}$. 
The dynamics $\bm F$ is then identified by formulating a sparse regression problem with a coefficient matrix $\bm{\Xi}$,
\begin{equation}
    \dot{\bm{H}}=\bm{\Phi}(\bm{\eta})\bm{\Xi},
    \label{eq:modeled_eq}
\end{equation}
with
\begin{equation}
    \bm{\Xi} = 
    \left[
    \begin{array}{cccc}
        \bm{\xi}_1 &  \bm{\xi}_2 & \dots & \bm{\xi}_{n_{\mathbb{\eta}}}
    \end{array}
    \right] = 
    \left[
    \begin{array}{cccc}
       \xi_{1,1}  & \xi_{2,1} &\dots& \xi_{n_{\mathbb{\eta}},1}\\
       \xi_{1,2}  & \xi_{2,2} &\dots& \xi_{n_{\mathbb{\eta}},2}\\
       \vdots     & \vdots    &\dots& \vdots\\
       \xi_{1,n_l} & \xi_{2,n_l} &\dots& \xi_{n_{\mathbb{\eta}},n_l}\\
    \end{array}
    \right],
    \label{coefficients}
\end{equation}
where $n_l$ is the number of library candidates.
We solve this sparse regression problem with the adaptive least absolute shrinkage and selection operator~\cite[adaptive LASSO;][]{Zou2006},
\begin{equation}
    \bm{\Xi}=\mathrm{argmin}||\bm{\dot{\eta}}-\bm{\bm{\Phi}(\bm{\eta})\bm{\Xi}}||^{2}+\alpha\sum_{j} w_{j}|\xi_{j}|
\end{equation}
where $\alpha$ denotes the weight to promote sparse coefficients.
We consider adaptive weights $w_{j}=(|\xi_{j}|)^{-\delta}$, which allows us to avoid the issue of multiple local minima in the optimization process~\cite{Fukami_Murata_Zhang_Fukagata_2021}.
In the present study, the identified ODEs are expected to capture the properties of the limit cycle oscillator and the perturbation dynamics in its vicinity. 
}

{
While the description above is provided with an arbitrary low-dimensional state vector ${\bm \eta}$, it is essential to choose appropriate low-order variables as ${\bm \eta}$ for the purpose of reconstructing spatial phase and amplitude sensitivity fields with a nonlinear decoder.
Based on our prior knowledge about sparse sensor reconstruction of flow fields~\cite{fukami2021,Zhong2023} and preliminary tests discussed in section~\ref{sec4}, we hereafter choose three sensor measurements placed on an airfoil surface, $\bm{s}\in\mathbb{R}^{3}$, as a low-order representation $\bm{\eta}$.
Hence, the modeled three-dimensional dynamics with sparse modeling for the sensor space is expressed as
\begin{equation}
\dot{\bm{s}}=\bm{\Phi}(\bm{s})\bm{\Xi},
    \label{eq:modeled_eq_sens}
\end{equation}
in a similar manner to equation~\ref{eq:modeled_eq}.
}

\subsection{Phase-amplitude reduction on low-order sensor space through adjoint analysis}
\label{phase reduction}

{
To characterize the limit cycle within the identified sensor space, phase-amplitude reduction is then introduced. 
Through this analysis, the perturbation dynamics can be analytically described with phase and amplitude, providing the mathematical foundation to derive the optimal control strategy for a post-stall airfoil wake.
}

{
Let us consider the dynamics within the three-dimensional sensor space, which exhibits a stable limit cycle, modeled in equation~\ref{eq:modeled_eq_sens}.
Corresponding to the periodic vortex shedding, the sensor dynamics feature a limit-cycle $\bm{s_0}(t)$ that satisfies $\bm{s_0}(t)=\bm{s_0}(t+T)$, where $T$ denotes the time period of the limit-cycle, and $\Omega_{n}=2\pi/T$ represents the corresponding natural frequency.
We assign a scalar phase to sensor variables defined as $\theta={\Theta}_{\bm s}(\bm{s_0})$ where $\Theta$ is the phase function. 
This phase function provides a linearized model of the original nonlinear dynamics on the limit cycle.
Extending the description to the vicinity of the limit cycle enables the description of perturbation dynamics and sensitivity analysis. 
Considering the sensor dynamics in the vicinity of the limit cycle, the phase $\theta$ monotonically increases at a constant rate as the natural frequency of the system $\Omega_{n}$, satisfying
\begin{equation}
    \dot{\theta}=\dot{\Theta}_{\bm s}(\bm{s})=\left\langle \nabla_{\bm s}\Theta_{\bm s}(\bm{s}), \dot{\bm{s}}\right\rangle = \left\langle \nabla_{\bm s}\Theta_{\bm s}(\bm{s}), \bm{F}(\bm{s})\right\rangle=\Omega_{n},
    \label{phase function}
\end{equation}
where the description of $\langle \bm{a},\bm{b}\rangle=\Sigma^{N}_{j=1}a^{*}_{j}b_j$ expresses the inner product, and $*$ denotes a complex conjugate.
}

{
Phase dynamics captures the effect of perturbation tangential to the limit cycle through phase advancement or delay. 
This is sufficient for frequency modification of periodic wakes.
However, solely phase-based control results in aerodynamic coefficients that are amplified, as the deviation from the limit cycle is not captured. In response, this study focuses not only on the speed of wake synchronization, but also aims to suppress the fluctuation of aerodynamic coefficients that are often amplified due to frequency modulation~\cite{Fukami2024, Godavarthi2023}.
We then further consider the amplitude degrees of freedom, which are described by $r_i={\bm R}_{{\bm s},i}(\bm{s}) (i=0,1,2)$.
The amplitude dynamics are characterized by the Floquet exponents $\lambda_0=0, \lambda_1,\lambda_2$, where the first term corresponds to the phase direction, and the others are sorted according to their real parts, as $\text{Re}(\lambda_0)=0\ge\text{Re}(\lambda_1)\ge\text{Re}(\lambda_2)$~\cite{Shirasaka2017,kuramoto2019}. 
This ordering implies that the first non-zero Floquet exponent $\lambda_1$ governs the slowest decaying amplitude mode. 
Therefore, the dominant term responsible for convergence toward the limit cycle is selected.
The dynamics then satisfies
\begin{equation}
    \dot{r}=\dot{R_{\bm s}}(\bm{s})=\left\langle \nabla_{\bm s} R_{\bm s}(\bm{s}), \dot{\bm{s}}\right\rangle = \left\langle \nabla_{\bm s} R_{\bm s}(\bm{s}), \bm{F}(\bm{s})\right\rangle=\lambda r(\bm{s}),
    \label{amplitude function}
\end{equation}
where $\lambda=\lambda_1$ is considered.
}

{
To account for the perturbation dynamics, let us then consider the influence of a weak external forcing on the reduced-order dynamics. 
The dynamics is described as
\begin{equation}
    \dot{\bm{s}}=\bm{F}(\bm{s})+\bm{f}(t),
\end{equation}
where the external perturbation $\bm{f}(t)$ is of order $ \mathcal{O}(\varepsilon)$ with $0 < \varepsilon \ll 1$. The dynamics of phase and amplitude with the perturbation are determined through
\begin{align}
&\dot{\theta}=\nabla_{\bm{s}}\Theta\cdot\dot{\bm{s}}=\nabla_{\bm{s}}\Theta\cdot[\bm{F}({{\bm s}_{0}})+\bm{f}(t)]=\Omega_{n}+\nabla_{\bm{s}}\Theta\cdot\bm{f},\\
&\dot{r}=\nabla_{\bm{s}}R\cdot\dot{\bm{s}}=\nabla_{\bm{s}}R\cdot[\bm{F}({{\bm s}_{0}})+{{\bm f}(t)}]=\lambda r+\nabla_{\bm{s}}R\cdot\bm{f}.
\end{align}
Assuming that the perturbation magnitude is small enough $\varepsilon\ll1$, these equations can be approximated by neglecting order $\mathcal{O}(\varepsilon^{2})$ terms, 
\begin{align}
    &\dot{\theta}=\Omega_{n}+\langle\bm{Z}_{\bm s}(\theta)\cdot\bm{f}(t)\rangle,\\
    &\dot{r}=\lambda r+\langle\bm{I}_{\bm s}(\theta)\cdot\bm{f}(t)\rangle,
\end{align}
where $\bm{Z}_{\bm s}(\theta)=\nabla_{\bm s}\Theta_{\bm s}|_{\bm{s}=\bm{s_0}}$ and $\bm{I}_{\bm s}(\theta)=\nabla_{\bm s} R_{\bm s}|_{\bm{s}=\bm{s_0}}$ denote the gradients of the phase and amplitude functions. These are the phase and amplitude sensitivity functions, respectively, as they determine the sensitivity of the dynamics with respect to the perturbation $\bm{f}(t)$.
These functions characterize the linear response in the phase-amplitude space to small perturbations, which are subsequently used to design the control strategy.
}

{
Although obtaining these sensitivity functions $\bm{Z}_{\bm s}(\theta)$ and $\bm{I}_{\bm s}(\theta)$ directly requires exact phase and amplitude functions $\Theta_{\bm s}$ and $R_{\bm s}$, evaluating them analytically remains challenging even within such a low-dimensional sensor space.
In response, an alternative approach based on the adjoint method is employed. Based on the linearized dynamics about the limit cycle, this method enables the efficient evaluation of these functions by solving the corresponding adjoint linear ordinary differential equations~\cite{Nakao2021},
\begin{align}
&    \frac{d}{dt}\bm{Z}_{\bm s}(t)=-\mathcal{J}^{\top}(t)\cdot\bm {Z}_{\bm s}(t),\label{phase_eq} \\
&    \frac{d}{dt}\bm{I}_{\bm s}(t)=-[\mathcal{J}^{\top}(t)-\lambda]\cdot \bm{I}_{\bm s}(t),
    \label{adjioint eq}
\end{align}
where the superscript $\top$ represents the transpose operation and $\mathcal{J}$ is a periodic Jacobian obtained from $\bm{F}$, modeled by SINDy. 
These sensitivity functions, $\bm{Z}_{\bm s}(\theta)$ and $\bm{I}_{\bm s}(\theta)$, describe the local sensitivity space of phase $\Theta_{\bm s}$ and amplitude $R_{\bm s}$ functions against perturbation in the sensor subspace, being extended to the global space of the flow field in the next section.
}

\subsection{Nonlinear decoder for reconstructing sensitivity fields}
\label{decoder}

{
We reconstruct the high-dimensional sensitivity fields from the low-order counterparts with nonlinear machine learning.
The full-field sensitivities can be computed through an adjoint solver of the Navier--Stokes equations~\cite{Kawamura2022}.
This, however, limits the applications of global field-based phase-amplitude reduction for experimental scenarios, where only a limited number of sparse sensor measurements are available.
Furthermore, it is challenging to accurately determine the phase and amplitude sensitivity functions, 
${\bm Z_{\bm q}}(\theta)=\nabla{\Theta_{\bm q}}|_{\bm{q}={\bm q}_0}$ and ${\bm I_{\bm q}}(\theta)=\nabla {R_{\bm q}}|_{\bm{q}={\bm q}_0}$, due to the high dimensionality of the flow field ${\bm q}$.
To bridge this gap, a nonlinear decoder is employed to reconstruct the sensitivity fields by using the sensitivity functions obtained in the sensor space. 
}

{
Let us consider a decoder $\mathcal{F}$ that accurately reconstructs the vorticity field $\bm{q}$ from the sparse sensor $\bm{s}$,
\begin{equation}
    \bm{q}\approx\hat{\bm{q}}=\mathcal{F}(\bm{s}),
    \label{eq:sparserec}
\end{equation}
where $\hat{\bm{q}}$ is an estimated vorticity field. 
The weights $\bm{w}$ inside the decoder are optimized by solving the minimization problem, 
\begin{equation}
    w^{*}=\underset{\bm{w}}{\text{argmin}}||\bm{q}-\mathcal{F}(\bm{s};\bm{w})||_{2}.
\end{equation}
This study considers two different models of the nonlinear decoder as ${\cal F}$, namely, a multi-layer perceptron (MLP) decoder~\cite{Rumelhart1986}, and a convolutional neural network-MLP (CNN-MLP) decoder~\cite{Lecun1998}, which will be discussed later.
}

{
Equipped with the decoder that estimates a high-dimensional vorticity field from sparse sensors, we have found that it is possible to leverage the local tangent space of the nonlinear decoder to approximate the high-dimensional sensitivity fields, ${\bm Z}_{\bm q}({\theta})$ and ${\bm I}_{\bm q}({\theta})$, from their low-dimensional counterparts, ${\bm Z}_{\bm s}({\theta})$ and ${\bm I}_{\bm s}({\theta})$.
Since phase and amplitude variables can be defined on the $C_l-\dot{C_l}$ plane, we posit that the high-dimensional flow field and the low-dimensional sensor space can share the identical phase and amplitude when the sensors are appropriately chosen such that
\begin{equation}
    \theta=\Theta_{\bm{q}}(\bm{q})=\Theta_{\bm{s}}(\bm{s}),
    \label{high-low phase}
\end{equation}
\begin{equation}
    r=R_{\bm{q}}(\bm{q})=R_{\bm{s}}(\bm{s}).
    \label{high-low amplitude}
\end{equation}
}

{
Since the phase and amplitude are now commonly shared between the high-dimensional field and the sensor space, differentiating equations~\ref{high-low phase} and~\ref{high-low amplitude} with respect to the sensor variables relates the two sensitivity functions through the multivariable chain rule.
Using the Jacobian $\bm{J}_{\mathcal{F}} = {\partial \mathcal{F}(\bm{s})}/{\partial \bm{s}} \in \mathbb{R}^{N_q \times N_{s}}$, which characterizes the local behavior of the decoder about the current sensor measurements, the sensor-space phase sensitivity is expressed as
\begin{align}
     \bm{Z}_{\bm{s}} = \frac{\partial \Theta_{s}}{\partial \bm{s}} = \left( \frac{\partial \bm{q}}{\partial \bm{s}} \right) ^{\top} \frac{\partial \Theta_{q}}{\partial \bm{q}} = \bm{J}_{\mathcal{F}}^{\top} \bm{Z}_{\bm{q}},
\end{align}
and likewise for the amplitude sensitivity $\bm{I}_{\bm{s}}=\bm{J}^{\top}_{\mathcal{F}}\bm{I}_{\bm{q}}$.
However, the decoder Jacobian $\bm{J}_{\mathcal{F}}$ is non-orthonormal, so the inverse mapping is no longer given by a simple transpose.
Moreover, because the number of sensors is significantly smaller than the full state dimension $(N_s \ll N_q)$, this can be a severely underdetermined system. 
We therefore compute the minimum-norm solution using the Moore--Penrose pseudo-inverse in the local tangent space, such that
\begin{align}
    \bm{Z}_{\bm{q}} \approx \bm{J}_{\mathcal{F}}(\bm{J}_{\mathcal{F}}^{\top} \bm{J}_{\mathcal{F}})^{-1}\bm{Z}_{\bm{s}},\label{pseudo inverse Z} \\
    \bm{I}_{\bm{q}} \approx \bm{J}_{\mathcal{F}}(\bm{J}_{\mathcal{F}}^{\top} \bm{J}_{\mathcal{F}})^{-1}\bm{I}_{\bm{s}}.
    \label{pseudo inverse I}
\end{align}
While pseudo-inverse operations for high-dimensional fields are typically prohibitively expensive, the use of equations \ref{pseudo inverse Z} and \ref{pseudo inverse I} can significantly reduce the computational cost in this sparse sensing framework, as the Gram matrix $\bm{J}_{\mathcal{F}}^{\top} \bm{J}_{\mathcal{F}}$ is only of size $N_s \times N_s$. 
Since the current number of sensors $N_s$ is very small ($N_s=3$), the computational cost of inverting this small matrix is practically negligible.
}

{
Because the sensitivity fields are reconstructed using an approximation rather than an exact adjoint-based derivation, they might be reliable only in the region near the sensors, whose uncertainty will be analyzed in section~\ref{sec4}.
However, since the optimal waveform is designed based on the optimal point of the approximated sensitivity fields, this approximation will be shown to be sufficient and effective for flow control, as also demonstrated in section~\ref{sec4}.
}

\subsection{Fast synchronization control with amplitude penalty}
\label{optimal entrainment}

{
This study considers frequency modification of aerodynamic flows through the lens of synchronization.
Let us consider the high-dimensional dynamics of the Navier--Stokes equation ${\cal N}$ with a periodic control input applied at a point $\bm{x}_f$ given by
\begin{equation}
    \frac{\partial\bm{q}}{\partial t}=\mathcal{N}(\bm{q}(\bm{x}, t))+\bm{b}(\bm{x},t),
    \label{NS eq controlled}
\end{equation}
where $\bm{b}(\bm{x}, t) = \bm{b}(\Omega_ft)\delta(\bm{x}-\bm{x}_f)$ is periodic with a forcing frequency $\Omega_{f}={2\pi}/ {T_f}$. 
Here, $T_f$ is the forcing period. Because the actuation is localized, the spatial coupling between the sensitivity field and the forcing reduces to their values at that point, and the phase dynamics under this external forcing becomes
\begin{equation}
    \dot{\theta}=\Omega_{n}+\bm{Z_q}(\theta)\cdot\bm{b}(\Omega_f t),
\end{equation}
where $\bm{Z_q}(\theta)$ hereafter denotes the reconstructed phase sensitivity field as described in section~\ref{decoder} evaluated at the actuation point $\bm{x}_f$.
}

{
To consider the synchronization between the system dynamics and the periodic forcing, we introduce the relative phase $\phi$ between the oscillator phase $\theta$ and the forcing phase $\Omega_{f}t$, as $\phi(t)=\theta(t)-\Omega_{f}t$. 
The relative phase dynamics then becomes
\begin{equation}
    \dot{\phi}(t)=\Delta+\bm{Z}_{\bm{q}}(\phi+\Omega_{f}t)\cdot \bm{b}(\Omega_{f}t),
\end{equation}
where $\Delta$ denotes frequency mismatch satisfying $\Delta=\Omega_{n}-\Omega_{f}$. 
Since this equation is non-autonomous, we perform temporal averaging~\cite{Kuramoto1984,hoppensteadt1997} over one period of input forcing resulting in,
\begin{equation}
    \dot{\phi}=\Delta+\Gamma(\phi),
    \label{relative phase eq}
\end{equation}
where
\begin{equation}
    \Gamma(\phi)=[\langle\bm{Z}_{\bm{q}}(\phi+\Omega_{f}s), \bm{b}(\Omega_{f}s)\rangle]_s
    \label{phase coupling function}
\end{equation}
is the phase coupling function and $[g(s)]_s=\frac{1}{T_f}\int^{T_f}_{0}g(s)ds$ represents the averaging over one forcing period. The synchronization is achieved when a given phase $\phi^{*}$ satisfies
\begin{equation}
    \Delta+\Gamma(\phi^{*})=0, \Gamma'(\phi^{*})<0.
\end{equation}
}

{
Once the relative phase converges to $\phi^{*}$, the oscillator phase is locked to $\theta=\Omega_ft+\phi^{*}$, and the frequency is locked to $\Omega_f$ on average. This phase locking occurs when $-\Gamma_{\text{max}}(\phi)<\Delta<-\Gamma_{\text{min}}(\phi)$ are satisfied. 
This condition describes the frequency range in which synchronization can be achieved.
}

{
Since we aim for fast frequency modification, we formulate an optimization problem to identify the optimal waveform that maximizes the synchronization speed. 
This can be expressed as,
\begin{equation}
    \begin{array}{c}
    \displaystyle
    \underset{\bm{b}}{\text{max}}~-\Gamma'(\phi^{*}),\\[16pt]
    \displaystyle
    \text{s.t.}\quad \Delta+\Gamma(\phi^{*})=0,\quad [\bm{b}\cdot\bm{b}]_t=P
    \end{array}
\end{equation}
where the input forcing power is constrained at a constant $P>0$. The cost function can be formulated as
\begin{equation}
    C(\bm{b}, \mu, \nu)=-\Gamma'(\phi^{*})+\mu\{\Delta+\Gamma(\phi^{*})\}+\nu(P-[\bm{b}\cdot\bm{b}]_t),
\end{equation}
where $\mu$ and $\nu$ are Lagrange multipliers.
The first term represents the synchronization speed, the second term constrains the synchronization condition, and the third term constrains the actuation power.
This optimization problem can be solved analytically using the calculus of variations~\cite{Zlotnik2013, Takata2021}. 
The optimal waveform is derived as
\begin{equation}
    \begin{array}{c}
    \displaystyle
    \bm{b}(\Omega_f t)=-\frac{1}{2\nu}{\bm Z}^\prime_{\bm q}(\phi^{*}+\Omega_f t)+\frac{\mu}{2\nu}{\bm Z}_{\bm q}(\phi^{*}+\Omega_f t),
    \end{array}
    \label{optimal waveform without amplitude penalty}
\end{equation}
where $\bm{Z}_{\bm{q}}'$ is the phase derivative of the phase sensitivity function. 
}

{
While this conventional optimization framework is designed to achieve fast synchronization solely by leveraging the phase sensitivity, it neglects the amplitude dynamics, which can lead to unintended deviation from the limit cycle.
Without accounting for amplitude dynamics, it may be detrimental to the aerodynamic performance, resulting in increased fluctuations even if fast synchronization is achieved.
To constrain such amplitude deviations, we introduce an amplitude penalty term into the objective function, formulated as
\begin{equation}
    \begin{array}{c}
    \displaystyle
    \underset{\bm{b}}{\text{max}}~-\Gamma'(\phi^{*})-k[(\bm{I}_{{\bm q}}(\phi^{*}+\Omega_{f}t)\cdot \bm{b}(\Omega_{f}t))^{2}]_{t},\\[16pt]
    \displaystyle
    \text{s.t.}\quad \Delta+\Gamma(\phi^{*})=0,\quad [\bm{b}\cdot\bm{b}]_t=P
    \end{array}
\end{equation}
where the coefficient $k\ge0$ denotes the penalty. 
The scalar product $\bm{I}_{{\bm q}}(\phi^{*}+\Omega_{f}t)\cdot \bm{b}(\Omega_{f}t)$ corresponds directly to the amplitude perturbation induced by the external perturbation in equation~\ref{amplitude function}.
By maximizing the second term, the influence of the forcing on the amplitude deviation from the baseline limit cycle is minimized while simultaneously achieving fast synchronization~\cite{Takata2021,Fukami2024}. 
The amplitude-penalized optimization employs the augmented cost function as
\begin{equation}
    D(\bm{b}, \mu, \nu)=-\Gamma'(\phi^{*})-k[(\bm{I}_{\bm q}(\phi^{*}+\Omega_{f}t)\cdot \bm{b}(\Omega_{f}t))^{2}]_t+\mu\{\Delta+\Gamma(\phi^{*})\}+\nu(P-[\bm{b}\cdot\bm{b}]_t),
\end{equation}
and the optimal waveform with amplitude penalty is hence derived as
\begin{equation}
\begin{aligned}
\bm{b}(\Omega_f t) &= \frac{1}{2}\left\{\nu\bm{E}+k \mathrm{Re}\left[ \bm{I}_{\bm q}(\phi^{*}+\Omega_{f}t)\bm{I}_{\bm q}^{\dagger}(\phi^{*}+\Omega_{f}t) \right] \right\}^{-1} \\
&\quad \times \left\{-\bm{Z}^\prime_{\bm q}(\phi^{*}+\Omega_{f}t)+\mu\bm{Z}_{\bm q}(\phi^{*}+\Omega_{f}t)\right\},
\end{aligned}
\label{optimal waveform with amplitude penalty}
\end{equation}
where $\bm{E}$ is an identity matrix. Although it is difficult to derive $\nu$ and $\mu$ analytically, the optimal waveform is obtained when we numerically determine the values of $\mu$ and $\nu$ by utilizing equation~\ref{optimal waveform with amplitude penalty}.
}

\section{Results}
\label{sec4}

\begin{figure}[t]
    \centering
    \includegraphics[width=1\textwidth]{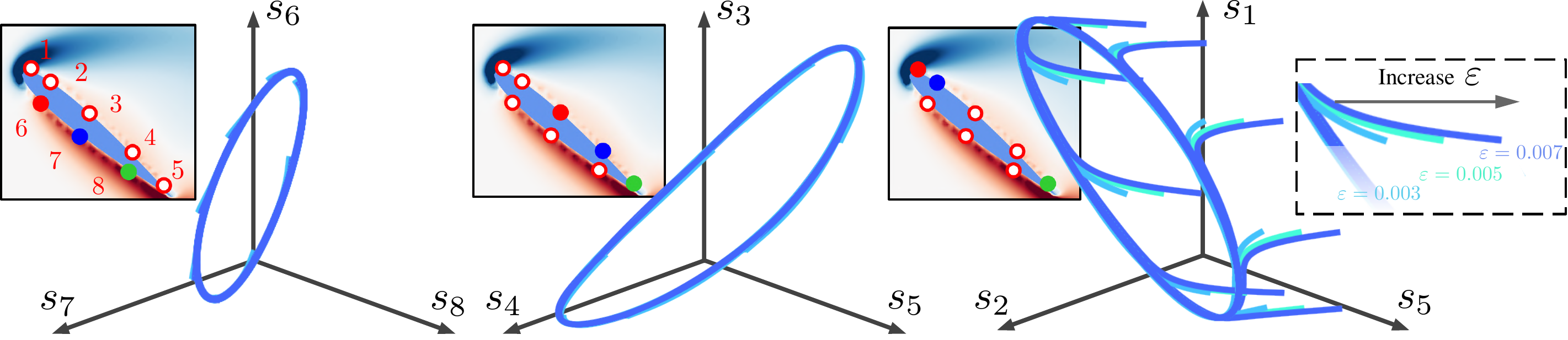}
    \caption{
    Three-dimensional subspace composed of different combinations of sparse sensors for an example case of NACA0012 airfoil at $\alpha=40^\circ$. 
    }
    \label{fig3}
\end{figure}

{
This section presents the high-dimensional phase and amplitude sensitivity fields reconstructed from three sensors, along with the fast manipulation of unsteady periodic flow fields via synchronization.
The validity of the reconstructed sensitivity fields is first evaluated. 
Leveraging these results, the proposed feedforward flow control with the amplitude constraint is then applied to achieve fast synchronization of unsteady aerodynamic flows.
}

{
The current approach introduced in section~\ref{sec3} can be applied to any low-order representations if there exists a decoder that accurately reconstructs high-dimensional vortical flow fields from the input low-order variables.
Before evaluating the performance of the reconstruction and control, let us discuss the rationale for selecting a sparse-sensor-based representation over other conventional reduced-order representations such as temporal coefficients of proper orthogonal decomposition. 
}

{
When defining the low-dimensional space, it is important that the chosen variables sufficiently capture the underlying physics.
In particular, the perturbation dynamics in the vicinity of the limit cycle need to be accommodated in the low-order representation to evaluate the phase and amplitude sensitivities.
In this study, these dynamics are observed by adding a pointwise impulse perturbation as a body force/jet. 
In this regard, we have found that sparse sensor measurements offer a superior candidate to other representations.
Our preliminary analysis has also found that such small perturbation effects cannot appear well in the coefficient space of proper orthogonal decomposition.
}

{
Furthermore, the spatial locations of the sensors are of significant importance.
To determine the optimal configuration, eight candidate sensors are distributed over the wing surface, measuring the local vorticity.
The resulting sensor spaces over trajectories for the different sensor configurations with an example case of NACA0012 airfoil at $\alpha = 40^\circ$ are illustrated in figure~\ref{fig3}.
While the first two trajectories shown there fail to capture the perturbation effects, the last configuration with sensors 1, 2, and 5 successfully captures the relevant dynamics.
Therefore, the present framework uses sensors 1, 2, and 5 for the cases across different thicknesses, at an angle of attack $\alpha=40^{\circ}$, whereas sensors 1, 3, and~5 are selected for the angles of attack $\alpha=50^{\circ}$ and $60^{\circ}$, following our similar preliminary investigations. 
}

{
For training the current data-driven models over different airfoil thicknesses and angles of attack, datasets with unperturbed and perturbed cases are prepared so that the model learns the dynamic behavior not only on the baseline limit cycle but also in its vicinity.
Each airfoil case contains $1000$ snapshots collected with a non-dimensional time step of $\Delta t=0.005$. To obtain the perturbed trajectories, the impulse actuation is added as a body force near the leading edge across equispaced eight different timings over one period as ${\theta=0, {\pi}/{4}, {\pi}/{2},...,{7\pi}/{4}}$. 
Three different actuation magnitudes of $\varepsilon=(0.003, 0.005, 0.007)$ are considered, as shown in figure~\ref{fig3}.
}

\begin{figure}
    \centering
    \includegraphics[width=1\linewidth]{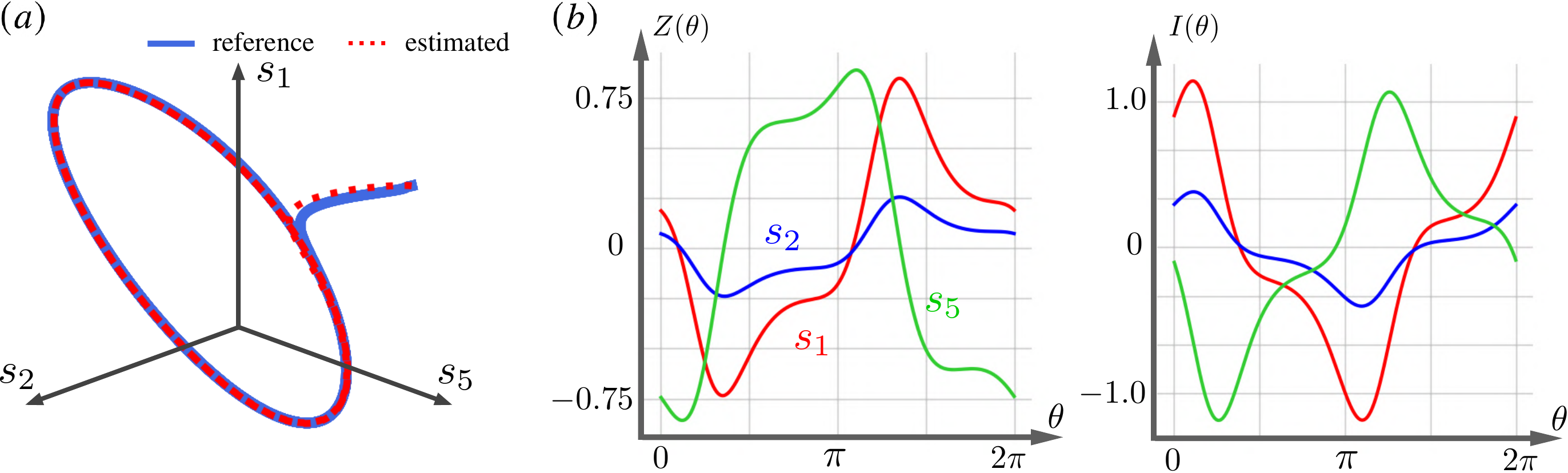}
    \caption{
    $(a)$ Comparison of reference and estimated traces in the sensor subspace for NACA0012 airfoil at $\alpha=40^\circ$. 
    $(b)$~Estimated phase and amplitude sensitivity functions.
    }
    \label{fig4}
\end{figure}

{
Let us discuss the sensitivity functions identified through the low-dimensional dynamical model in the sensor space.
To examine whether the identified model correctly captures the physical trends, the integrated trajectory over time is compared with the reference trajectory, as shown with an example case of NACA0012 airfoil at $\alpha = 40^\circ$ in figure~\ref{fig4}$(a)$. 
Following the impulsive perturbation, the trajectory deviates from the limit cycle, subsequently returns to the limit cycle, and after a sufficient duration, aligns with the dynamic behavior in the high-dimensional flow field.
}

{
The phase and amplitude sensitivity functions are then estimated by solving the adjoint linear ordinary differential equations \ref{phase_eq} and \ref{adjioint eq}, as exhibited in figure~\ref{fig4}$(b)$. 
For this calculation, the Jacobian matrix evaluated from the SINDy modeling is used.
The resulting sensitivity functions exhibit $2\pi$ periodicity with maximum and minimum peaks near $\theta=0$ and $\theta=\pi$. 
The absolute values differ among the three sensors $s_{1}, s_{2}$ and $s_5$.
This variance might be attributed to the spatial locations of the sensors. 
Since sensors $s_{1}$ and $s_5$ are located near the leading and trailing edges, respectively, they are likely more sensitive, as the perturbation directly affects the vortex shedding process~\cite{Eldredge2018,Godavarthi2023}. 
}

{
The amplitude sensitivity function achieves maximum peaks at different times than the phase sensitivity function. 
This indicates that the influence of perturbation is different in both the phase and amplitude directions and gives an additional degree of freedom to control the amplitude deviation.
Note that the sensitivity functions obtained here represent the relative sensitivity within the sparse sensor space, rather than the exact, absolute sensitivity values in the high-dimensional flow field.
}

\begin{figure}
    \centering
    \includegraphics[width=1\linewidth]{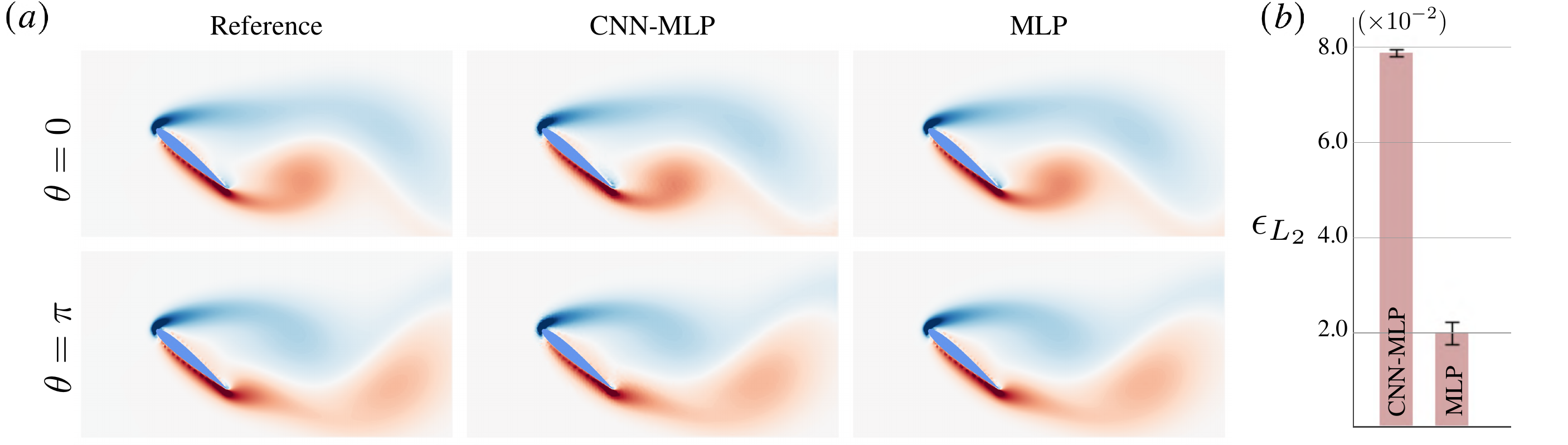}
    \caption{
    $(a)$ Reconstructed vorticity flow fields of the NACA0012 case at $\alpha = 40^\circ$
    using the CNN-MLP and the MLP models for equation \ref{eq:sparserec}.
    $(b)$ The ensemble $L_2$ error norm~$\epsilon_{L_2}$ over the test snapshots.
    }
    \label{fig5}
\end{figure}

{
As previously noted, to properly apply the phase-amplitude reduction analysis, accurately capturing the dynamics in the vicinity of the limit cycle via not only the low-order model but also the decoder is needed. 
To achieve accurate full-field reconstruction, this study compares the MLP decoder~\cite{erichson2020,FFT2020} and the CNN-MLP decoder~\cite{Fukami_Byungin_2022}, often used for sparse sensor reconstruction of unsteady flows, as shown in figure~\ref{fig5}. 
}

{
While both the MLP decoder and the CNN-MLP decoder accurately reconstruct vortical flows from three sensors, the MLP one reports an approximately four times smaller $L_2$ reconstruction error norm.
Here, the $L_2$ error norm is defined as $\epsilon_{L_2} = ||{\bm q}-{\cal F}({\bm s})||_2/||{\bm q}||_2$.
Although the training cost of the MLP model is a bit higher than that of the CNN-MLP model due to its fully-connected nature~\cite{chen2024sparse}, the MLP decoder exhibits a superior capability in reconstructing the transient process from the perturbation to the unperturbed state.
Therefore, the MLP decoder is selected for subsequent analysis.
}

{
Next, the spatial reliability of the sensitivity field is investigated.
Theoretically, the reconstructed sensitivity fields are independent of the sensor choice if enough sensors and a linear decoder are used.
However, since the reconstructed spatial sensitivity fields are obtained through the nonlinear decoder, their reliability can be determined by varying the sensor location.
As discussed in section~\ref{decoder}, the sensitivity fields, ${\bm Z}_{\bm q}$ and ${\bm I}_{\bm q}$, are an approximation through the nonlinear decoder.
We posit that, with a given sensor-reading placement shown in figure~\ref{fig3}, the reconstructed sensitivity fields $\bm{Z}_{\bm{q}}$ and $\bm{I}_{\bm{q}}$ through \ref{pseudo inverse Z} and \ref{pseudo inverse I}, do not vary significantly under a slight shift of sensor locations.
In other words, if the sensitivity fields remain consistent across different sensor configurations with slight modifications, the corresponding spatial region might be deemed reliable.
}

\begin{figure}
    \centering
    \includegraphics[width=1\linewidth]{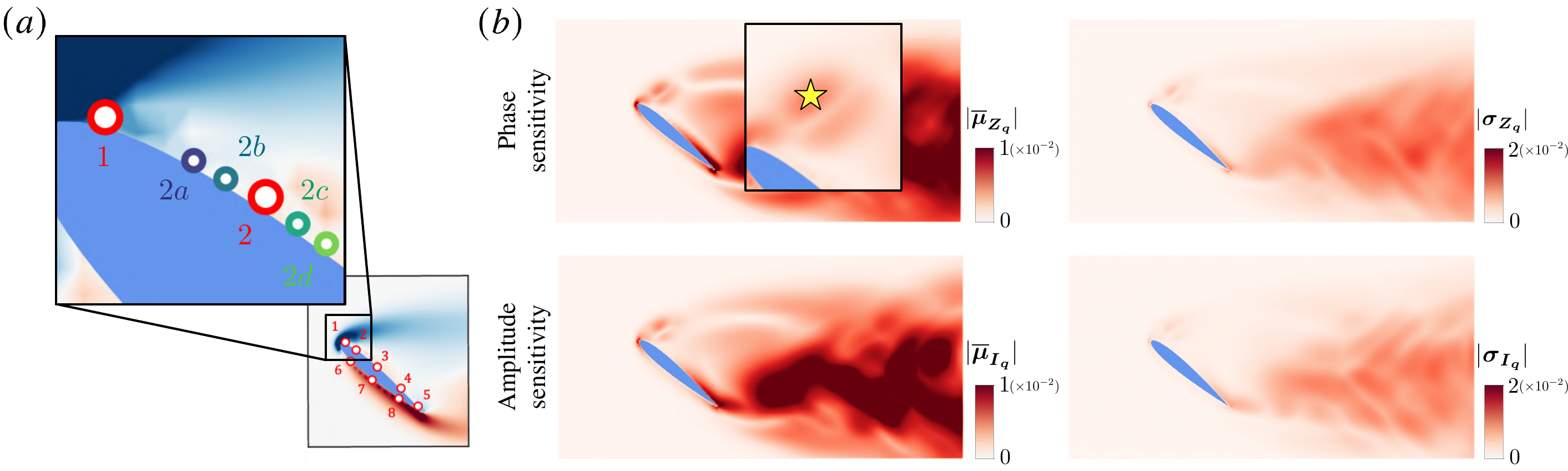}
    \caption{Spatial reliability of reconstructed sensitivity fields for NACA0012 at $\alpha=40^{\circ}$
    $(a)$ Sensor distribution for evaluating sensitivity field uncertainty.
    $(b)$ Mean and standard deviation of the phase and amplitude sensitivity fields.
    }
    \label{fig6}
\end{figure}

{
With an example of a NACA0012 airfoil at $\alpha = 40^\circ$, we examine the uncertainty of the resulting sensitivity fields by sweeping the location of $s_2$ near the base locations, denoted as ${2_{a}, 2_{b}, 2_{c}, 2_{d}}$, while the other two sensors 1 and 5 remain fixed, as illustrated in figure~\ref{fig6}$(a)$.
To assess the reliability, we compute the time-ensemble average $\overline{\bm{\mu}}_{\bm Z_{\bm q}}$ and the corresponding standard deviation field ${\bm \sigma}_{\bm{Z}_{\bm q}}$ over~$n_t$ snapshots as
\begin{equation}
\overline{\bm{\mu}}_{\bm Z_{\bm q}} = \frac{1}{n_sn_t} \sum_{t=1}^{n_t} \sum_{i=1}^{n_s} \bm{Z}_{{\bm q},i,t}, \quad \bm{\sigma}_{\bm{Z}_{\bm q}} = \sqrt{\frac{1}{n_sn_t-1} \sum_{t=1}^{n_t} \sum_{i=1}^{n_s} \left( \bm{Z}_{{\bm q},i,t} - \frac{1}{n_s}\sum_{i=1}^{n_s}\bm{Z}_{\bm{q}, i, t} \right)^{2}}
\end{equation}
where index $i$ denotes the sensor location.
The number of sensor locations swept for this analysis is set to be $n_s = 5$, as mentioned above.
The descriptions for the amplitude sensitivity are omitted for brevity.
}

{
The calculated sensitivity fields are exhibited in figure~\ref{fig6}$(b)$.
The mean sensitivity field exhibits a consistent spatial structure across varying sensor locations, whereas the standard deviation fields represent the spatial distribution of uncertainty.
The wake region in the standard deviation exhibits high uncertainty in both the phase and amplitude sensitivity fields.
The region near the airfoil, particularly around the leading edge, presents reasonable reliability. 
}

{
This localized certainty is partly attributed to the fact that the impulse actuation is applied near the leading edge. 
As physical actuation is typically executed near the wing in a practical sense, we argue that these accurate sensitivity distributions near the leading edge would be useful from the aspect of flow modification.
We will show how the current sensitivity can be used to quickly control the wake shedding frequency while suppressing the amplitude modulation.
Within the reliable leading edge region, the ensemble-averaged sensitivity field exhibits a distinct maximum sensitive point, marked with a star symbol, which will subsequently be selected as the target point for extracting the sensitivity function for flow control later.
Note that a higher number of actuation points might be needed to obtain sensitivity fields with less uncertainty near the wake region.
}

\begin{figure}
    \centering
    \includegraphics[width=0.9\linewidth]{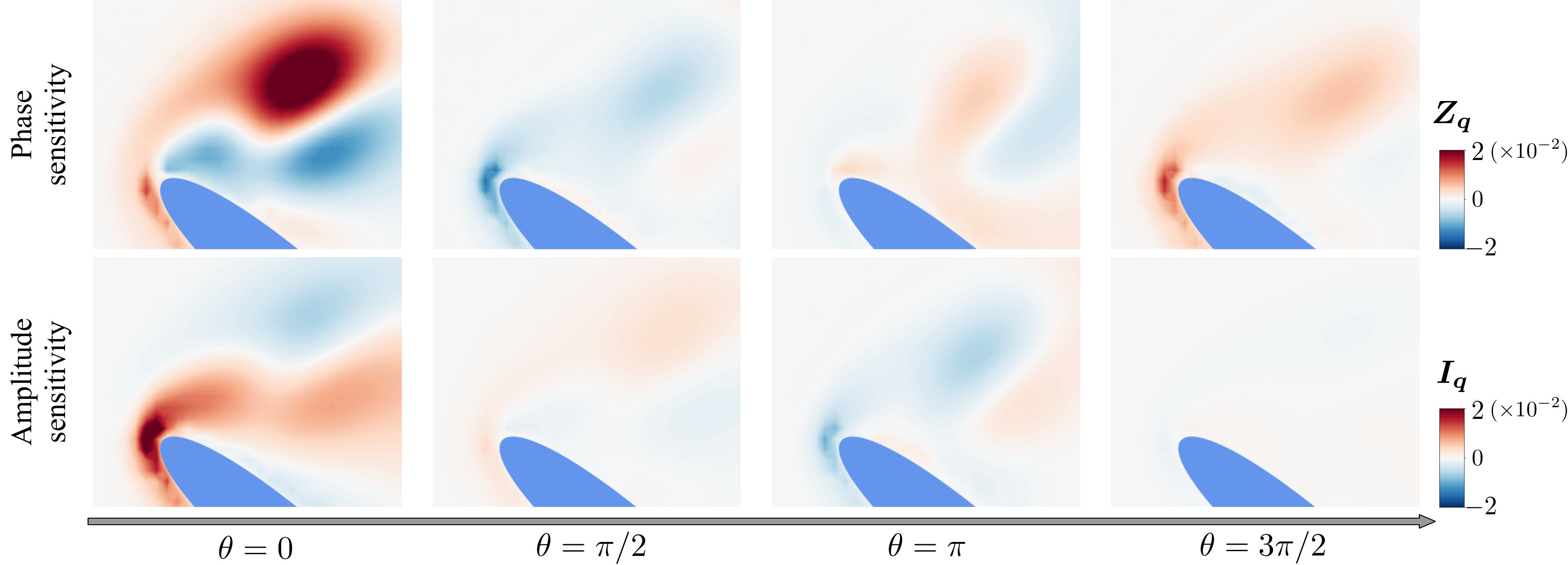}
    \caption{
    Reconstructed phase and amplitude sensitivity fields, ${\bm Z}_{\bm q}(\theta)$ and ${\bm I}_{\bm q}(\theta)$ over a period of the NACA0012 airfoil at $\alpha = 40^\circ$.
    }
    \label{fig7}
\end{figure}

{
As the sensitivity fields are at hand with the present framework, let us then examine the dependence of the phase and amplitude sensitivity fields on the angle of attack and airfoil thickness. 
The representative case with $\alpha = 40^{\circ}$ for NACA0012 airfoil is presented in figure~\ref{fig7}. 
We observe maximum phase sensitivity at $\theta=0$ for the phase and amplitude sensitivity fields in these four time windows. 
Applying a positive vortical perturbation at a red region in the phase sensitivity field directly induces phase advancement, and conversely, a perturbation applied at a blue region results in phase delay.
A similar positive perturbation applied to a red region in the amplitude sensitivity field leads to an increase in fluctuation from the baseline limit cycle.
These observations suggest that na\"ively modifying the frequency based solely on the phase sensitivity fields likely causes unexpected excitation of amplitude dynamics.
}

\begin{figure}
    \centering
    \includegraphics[width=0.9\linewidth]{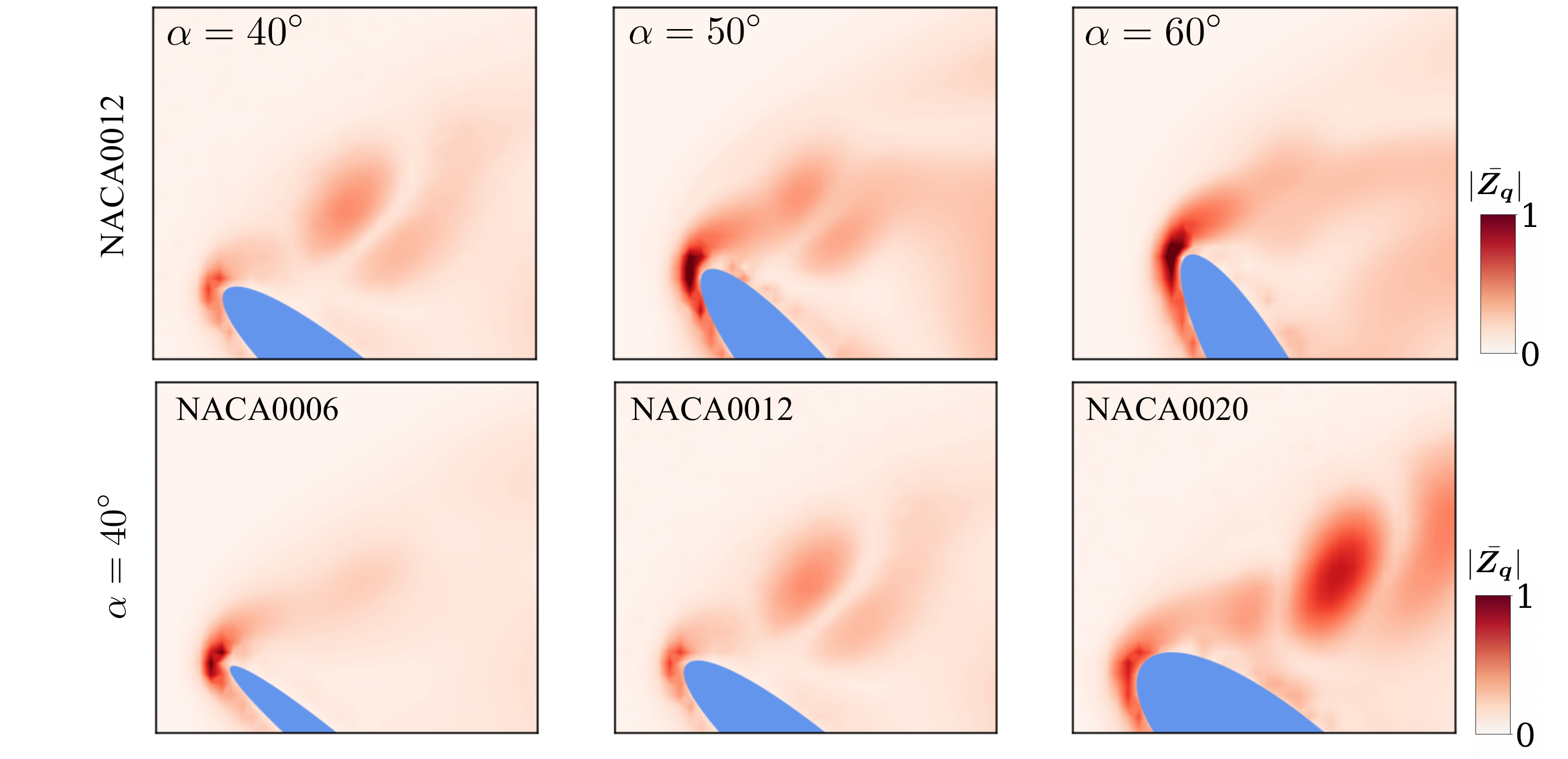}
    \caption{
    Comparison of time-averaged absolute sensitivity fields across various airfoil parameters.
    }
    \label{fig8}
\end{figure}

{
Let us then examine the influence of varying the angle of attack and airfoil thickness. 
Since the sensor location is adjusted as the angle of attack varies, the sensitivity fields are evaluated using the absolute value of the time-averaged fields, as shown in figure~\ref{fig8}. 
As $\alpha$ increases, the spatial extent of the intensely sensitive region contracts. 
This contraction occurs because an increase in the angle of attack enhances earlier vortex separation and intensifies the unsteadiness of the wake~\cite{fukami_taira_2023}.
This requires a larger amount of actuation to modify the vortex shedding process at a higher angle of attack.
Meanwhile, the sensitive region intensifies and expands as the airfoil thickness increases. 
This expansion is attributed to the lower curvature of a thicker airfoil, which delays the vortex detachment near the leading edge and mitigates the flow unsteadiness, leading to a broader sensitivity distribution.
The trend in the sensitivity fields coincides with a previous study discussing the adjoint-based phase-reduction analysis of airfoil wakes~\cite{Kawamura2022}.
}

{
Equipped with the sensitivity fields, let us then perform fast optimal entrainment of the unsteady periodic flow.
We consider a post-stall flow over the NACA0012 airfoil at an angle of attack $\alpha=40^{\circ}$. 
We compare the control performance across three distinct forcing configurations: sinusoidal forcing~\cite{Taira2018}, the fast optimal forcing without constraint on the amplitude deviation~\cite{Godavarthi2023}, and the fast optimal forcing with the amplitude penalty.
The current frequency-based flow modification is evaluated in terms of the speed of synchronization, the lift coefficient, and the vorticity field.
}

\begin{figure}
    \centering
    \includegraphics[width=1\linewidth]{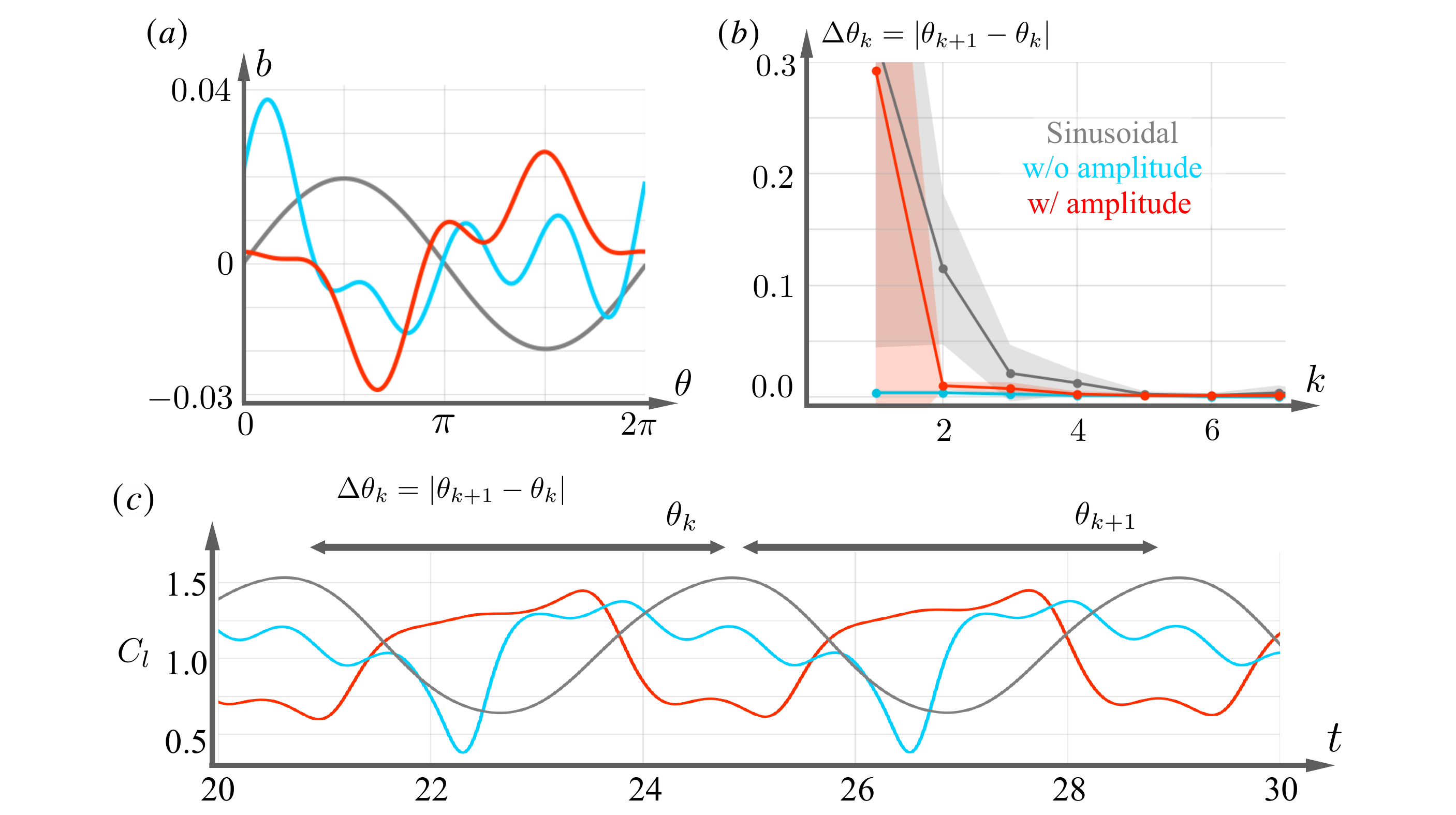}
    \caption{
    Rapid flow modification for periodic airfoil wakes.
    $(a)$ Forcing configurations considered in this study.
    $(b)$ Synchronization speed assessed via the inter-peak phase difference $\Delta \theta_k$.
    We show uncertainties in $\Delta \theta$ as a shade plot for each waveform, obtained by perturbing the flow with eight equispaced initial conditions over the phase.
    $(c)$ Time series of the lift coefficient $C_l$.
    }
    \label{fig9}
\end{figure}

{
For this design, the phase and amplitude sensitivity functions extracted from the maximum sensitivity point identified in section~\ref{sec3} are used. 
The optimal waveforms are designed via equations~\ref{optimal waveform without amplitude penalty} and~\ref{optimal waveform with amplitude penalty} using the phase and amplitude sensitivity functions sampled at the chosen actuation location, as shown in figure~\ref{fig9}$(a)$.
The forcing frequency is set to $\Omega_{f}=1.05\Omega_{n}$ to achieve faster synchronization relative to the shedding frequency $\Omega_n$.
For comparison, a sinusoidal waveform forcing is also considered.
}

{
The optimal waveform without amplitude penalty exhibits a single peak around $\theta=\pi/8$, representing the critical actuation phases required to achieve quick synchronization.
In contrast, the optimal waveform with amplitude penalty exhibits two peaks around $\theta=3\pi/4$ and $\theta=3\pi/2$. 
This difference suggests that while the phase position at $\theta=\pi/8$ is inherently effective for modification of phase dynamics, actuation at $\theta=\pi/8$ induces unintended amplitude deviations. 
In turn, actuation at $\theta=3\pi/4$ and $\theta=3\pi/2$ achieves the target frequency synchronization without too much amplitude growth, which coincides with the observation in figure~\ref{fig7}.
}

{
We introduce actuation at the location near the leading edge with maximum sensitivity, as shown in figure~\ref{fig6}.
The periodic forcing based on the considered waveforms is added as a body force discussed in section~\ref{optimal entrainment} at a forcing frequency of $\Omega_f=1.05\Omega_n$.
To compare the synchronization speed, we consider the inter-peak phase difference,
\begin{align}
\Delta\theta_{k} = |\theta_{k+1} - \theta_k|,    
\end{align}
where $k$ denotes the index for each cycle extracted from the lift coefficient, as shown in figures~\ref{fig9}$(b)$ and $(c)$, following Godavarthi $\it{et}$ $\it{al.}$ \cite{Godavarthi2023}. 
}

{
Phase synchronization is successfully achieved when this inter-peak difference converges to zero~\cite{Godavarthi2023}.
To account for the uncertainty associated with the initial actuation timing, the inter-peak differences are examined by introducing actuation at $\theta=0, \pi/2, \pi, 3\pi/2, 2\pi$. 
The convergence speed to synchronization is evaluated in an ensemble-averaged manner in figure~\ref{fig9}$(b)$, in which each solid line corresponding to the actuation waveform represents the mean and the shaded region denotes $95\%$ confidence intervals to account for dependence on the initial actuation phase.
The optimal waveforms achieve synchronization much faster than the sinusoidal forcing because of the assistance of the phase sensitivity function. 
These results demonstrate the validity and efficacy of the approximated phase sensitivity fields reconstructed only from three sparse sensors.
}

{
Let us then focus on the amplitude modulation.
The optimal waveform with the amplitude penalty is motivated to minimize deviations from the limit-cycle trajectory while achieving fast synchronization. 
This enables special care of aerodynamic coefficients, such as the decrement of the lift coefficient.
The time trace of the lift coefficient $C_l$ is evaluated in figure~\ref{fig9}$(c)$.
}

{
To quantitatively assess the amplitude deviations, the statistics of the lift coefficient of the controlled flow are summarized in table~\ref{tab}.
The range between the minimum and maximum values quantifies the magnitude of the amplitude deviation.
The optimal waveform with amplitude penalty successfully achieves an approximate $20\%$ mitigation of the amplitude deviation of the lift coefficient compared to the phase-only based control waveform.
}

\begin{figure}
    \centering
    \includegraphics[width=1\linewidth]{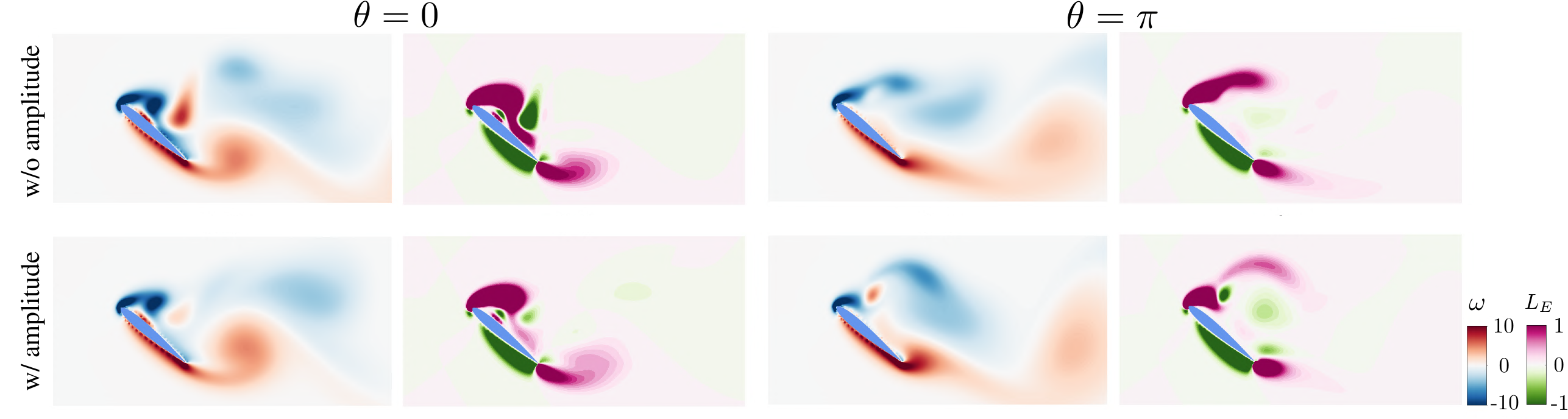}
    \caption{
    Instantaneous lift force elements $L_E$ at different forcing configurations
    }
    \vspace{-4mm}
    \label{fig10}
\end{figure}

{
To further examine the effect of the current forcing, the lift-element fields~\cite{chang1992} are compared in figure~\ref{fig10}.
The lift-element analysis enables the identification of the vortex contribution to the lift generation by decomposing the lift force.
For decomposition, an auxiliary potential function $\phi_L$ is necessary at every time step, satisfying the Laplace equation $\nabla^{2}\phi_L=0$, and the boundary condition $\nabla\phi_L \cdot \bm{n}=-\bm{n}\cdot \bm{e}_y$ on the airfoil surface, where $\bm{n}$ is the unit vector normal to the airfoil and $e_y$ is the unit vector in the vertical direction.
The total lift force is decomposed as  
\begin{align}
    F_L =  \int_V (\boldsymbol{\omega} \times \mathbf{u} ) \cdot \nabla \phi_L \, dV + \frac{1}{Re} \int_S (\boldsymbol{\omega} \times \mathbf{n}) \cdot \left( \nabla \phi_L + \mathbf{e}_y \right) \, dS,
\end{align}
where the first term integrand on the right-hand side represents the lift element $L_E$, and the second term denotes the surface vorticity contribution to the lift force.
}

{
The lift element field in the case without amplitude penalty at $\theta=0$ exhibits a larger-scale vortex structure behind the leading-edge vortex than in the case with amplitude penalty. 
At $\theta=\pi$, on the other hand, the case with amplitude penalty introduces a secondary vortex structure that breaks down the large leading-edge vortex.
This is also reflected as a negative lift element downstream of the leading-edge vortex, while the case without penalty shows a larger positive lift element near the leading edge.
This timing is effective for modifying the phase dynamics while remaining insensitive to the amplitude dynamics, thereby reducing the lift fluctuation.
These results can support the fidelity and appropriateness of the estimated amplitude sensitivity fields evaluated by the present technique. 
}

\section{Concluding remarks}
\label{sec5}

{
We presented a fast frequency modification of the periodic airfoil wake using phase-amplitude reduction and nonlinear machine learning-based sparse sensor reconstruction.
We considered post-stall unsteady periodic wakes of NACA airfoils at a chord-based Reynolds number of $Re=100$ across various geometric parameters.
To derive the sensitivity fields of phase and amplitude variables, a data-driven framework integrating the sparse identification of nonlinear dynamics, the adjoint method, and a nonlinear decoder is proposed.
With mere three sensors on the wing, we approximated phase and amplitude sensitivity fields that otherwise require a numerical adjoint solver.  
Through the approximated phase and amplitude sensitivity fields, we identified an optimal actuation location at the leading edge. 
We derived the actuation waveforms optimized for fast synchronization to the forcing frequency with and without constraining the amplitude deviation.
The current control strategy outperforms the sinusoidal waveform in terms of flow modification speed, while successfully bounding the amplitude deviations. 
Consequently, these analyses validate the efficacy of the sensor-based sensitivity field reconstruction.
}

{
The current technique achieves fast synchronization-based control with the suppression of amplitude deviation actuation, which avoids the degradation of aerodynamic performance. 
The present approach reconstructs the phase sensitivity field near the leading edge around the perturbed location, which is acceptable from the perspective of practical flow modification.
While deriving the full-field sensitivity has often been achieved with the adjoint solver~\cite{Kawamura2022}, which is often computationally expensive for identifying the sensitivity fields, the current data-driven technique can potentially be applied to experimental scenarios in the absence of the numerical solver.
Furthermore, the proposed technique provides the amplitude sensitivity fields, thereby enabling amplitude-constrained control while facilitating the understanding of flow physics.
}

{
Although the present study focused on low-Reynolds-number flows around an airfoil, extending this framework to more complex, three-dimensional flows at higher Reynolds numbers that can be modeled as multi-frequency oscillators remains of interest for future research.
Furthermore, because this framework relies primarily on data availability rather than field-specific governing equations, it may be applicable to various other research domains beyond aerodynamics, including chemical or biological systems exhibiting periodic or rhythmic dynamics. 
}

\begin{table}
  \begin{center}
  \centering
  \noindent\rule{\textwidth}{0.4pt}
  \vspace{2pt}
  
  \def~{\hphantom{0}}
  \begin{tabular}{lcccc}
    Case                       &    max   &   min  &  max-min  &   ~mean  \\[3pt]
    No control                    &  1.115   & ~0.849 &  0.266  &  ~0.975  \\
    Sinusoidal wave             &  1.532   & ~0.643 &  0.889  &  ~1.078  \\
    Optimal wave w/o amplitude  &  1.381   & ~0.382 &  0.999  &  ~1.059  \\
    Optimal wave w/ amplitude   &  1.449   & ~0.631 &  0.818  &  ~1.085  \\
  \end{tabular}
  \vspace{12pt}
  \caption{
  Comparison of the maximum, minimum, and mean values of the lift coefficient~$C_l$ for different control strategies corresponding to the NACA0012 airfoil at $\alpha=40^\circ$.
  }
  \label{tab}
  \vspace{2pt}
  \noindent\rule{\textwidth}{0.4pt}
  \end{center}
\end{table}

\section*{Acknowledgments}{
{
K.F. acknowledges support from 
the JSPS KAKENHI Grant No.~JP25K23418 and No.~JP26K01129,
the JST PRESTO Grant No.~JPMJPR25KA, 
and the MEXT Coordination Funds for Promoting Aerospace Utilization Grant No.~JPJ000959.
H.N. acknowledges support from the JSPS KAKENHI Grant No.~JP25H01468, and No.~JP25K03081.\\
}} 

\section*{Declaration of interests}{The authors report no conflict of interest.}


\bibliographystyle{unsrt}  
\bibliography{refs}

\end{document}